# ALMA Discovery of Solar Umbral Brightness Enhancement at λ=3 mm

**Running Head**: Umbral brightness enhancement at λ=3 mm

Kazumasa Iwai[1*], Maria Loukitcheva[2, 3, 4], Masumi Shimojo[5], Sami K. Solanki[3, 6] and Stephen M. White[7]

1. National Institute of Information and Communications Technology, 4-2-1 Nukui-kita, Koganei, Tokyo 184-8795, Japan
2. Center for Solar-Terrestrial Research, New Jersey Institute of Technology, 323 Martin Luther King Boulevard, Newark, NJ 07102, USA
3. Max Planck Institute for Solar System Research, Justus-von-Liebig-Weg 3, 37073 Göttingen, Germany
4. Astronomical Institute, St. Petersburg University, Universitetskii pr. 28, 198504 St. Petersburg, Russia
5. Chile Observatory, National Astronomical Observatory of Japan, Mitaka, Tokyo 181-8588, Japan
6. School of Space Research, Kyung Hee University, Yongin, Gyeonggi 446-701, Republic of Korea
7. Space Vehicles Directorate, Air Force Research Laboratory, Albuquerque, NM, USA

*Corresponding author email: kazumasa.iwai@nict.go.jp

## Abstract

We report the discovery of a brightness enhancement in the center of a large sunspot umbra at a wavelength of 3 mm using the Atacama Large Millimeter/sub-millimeter Array (ALMA). Sunspots are amongst the most prominent features on the solar surface, but many of their aspects are surprisingly poorly understood. We analyzed a λ=3 mm (100 GHz) mosaic image obtained by ALMA, which includes a large sunspot within the active region AR12470 on December 16, 2015. The 3 mm map has a 300"×300" field-of-view and 4".9×2".2 spatial resolution, which is the highest spatial-resolution map of an entire sunspot in this frequency range. We find a gradient of 3 mm brightness from a high value in the outer penumbra to a low value in the inner penumbra/outer umbra. Within the inner umbra, there is a marked increase in 3mm brightness temperature, which we call an umbral brightness enhancement. This enhanced emission corresponds to a temperature excess of 800 K relative to the surrounding inner penumbral region and coincides with excess brightness in the 1330 and 1400 Å slitjaw images of the Interface Region Imaging Spectrograph (IRIS), adjacent to a partial lightbridge. This λ=3 mm brightness enhancement may be an intrinsic feature of the sunspot umbra at chromospheric heights, such as a manifestation of umbral flashes, or it could be related to a coronal plume since the brightness enhancement was coincident with the footpoint of a coronal loop observed at 171 Å.

**Key words:** sunspots – Sun: chromosphere – Sun: radio radiation

1. Introduction

Sunspots and in particular their umbrae are darker than their surroundings at optical wavelengths because the strong magnetic field of sunspots inhibits convective energy transport, leading to a decrease in temperature of the gas in the sunspot, at least in photospheric layers (Rempel and Schlichenmaier 2011). In contrast, sunspots can be neutral or even appear bright in radiation emanating from higher layers of the atmosphere, such as the chromosphere. This is because of excess heating in those layers that is presumably also related to the strong magnetic field, although the exact cause and even the magnitude of the heating are poorly known (Solanki 2003).

To determine how and how much energy flows through the solar atmosphere, we need a profile of the temperature and density as a function of height. There have been many such models, and most of these are based on strong atomic lines in the ultraviolet (UV) and infrared range. However, these lines are thought to be formed under non-local thermodynamic equilibrium (non-LTE) conditions. Hence, the observational results require non-LTE radiative-transfer simulations to facilitate their interpretation. In fact, available atmospheric models for sunspot umbrae show a large scatter of the predicted temperature at the chromosphere (see Figure 6 of Loukitcheva et al. 2014).

Solar radio emission at millimeter wavelengths arises in the chromosphere and can be used to constrain atmospheric models (e.g. Vernazza et al. 1978). The opacity of the thermal free-free emission at millimeter and submillimeter wavelengths depends on the emission measure and temperature profiles in the atmosphere (Dulk 1985). In addition, the Rayleigh-Jeans law and LTE conditions are applicable for these wavelengths. Hence, we can estimate the radio brightness temperature expected from vertical distributions of the atmospheric density and temperature in a given model.

There have been several observational studies of sunspots at submillimeter and millimeter wavelengths. In the submillimeter range, sunspot observations using the Caltech Sub-millimeter Observatory (CSO) by Bastian et al (1993), and the James Clerk Maxwell telescope (JCMT) by Lindsey et al (1995), suggested that the brightness temperature of a sunspot umbra is lower than that of quiet Sun regions. At a wavelength of 3 mm, White et al (2006) used the Berkeley–Illinois–Maryland Array (BIMA) interferometer, and Iwai and Shimojo (2015) have used the Nobeyama 45m telescope to measure the brightness temperature of sunspots. In both cases, the umbra was darker than the surrounding atmosphere, albeit with a spatial resolution of 12-16 arcsec that limited their ability to separate typical sunspot umbrae from the surrounding bright penumbra and active region. At a wavelength of 8.8 mm, Iwai et al (2016) found that sunspots are generally not distinguishable from the surrounding atmosphere using the Nobeyama Radioheliograph (NoRH).

The main problem with previous observations is the relatively poor spatial resolution, which is insufficient to cleanly resolve the umbra from the penumbra and the surrounding atmosphere, or to identify any sub-structure within the umbra or penumbra. The inability to resolve the different structures limits the ability to infer the actual umbral temperature from the observations.

The high spatial resolution of the Atacama Large Millimeter/sub-millimeter Array (ALMA; Hills, et al. 2010) in the millimeter and sub-millimeter range is sufficient to well-resolve the umbra of a typical sunspot. The purpose of this study is to explore the structure of a sunspot at $\lambda$=3 mm in high-resolution ALMA images and to compare it with ultraviolet (UV) and extreme ultraviolet (EUV) images obtained by the Interface Region Imaging Spectrograph (IRIS) and the Solar Dynamics Observatory (SDO). The instrument and data set used in this study are described in Section 2. The data analysis is presented in Section 3. The results are summarized and discussed in Section 4.

2. Observation

The observations discussed here were carried out with ALMA during solar commissioning activities between 18:01 and 18:48 on 2015 December 16 and released as ALMA Science Verification data in early 2017. The Band 3 ($\lambda$=3 mm, i.e. 100 GHz) observation was carried out in a compact array configuration which includes twenty-two 12m antennas and nine 7m antennas. Visibilities obtained from both 12m and 7m antennas and their combinations were included in the image synthesis. The full-width at half-maximum (FWHM) of the resulting synthesized beam (spatial resolution) is 4".9×2".2. The map is derived from a mosaic observation of 149 pointings to cover the field-of-view of 300"×300". Details of the observation and analysis are provided by Shimojo et al. (2017).

Single dish, fast-scanning observations covering the full solar disk were carried out simultaneously (see White et al., 2017). The FWHM of the single-dish primary beam is about 60" at λ=3 mm. The single dish and interferometric data were combined in the U-V plane (feathering) to derive the absolute brightness temperature of the interferometric maps (Shimojo et al. 2017).

The receivers of both interferometric and single dish antennas were de-tuned for the mixer mode 2 (MD2) setting (Shimojo et al, 2017) to prevent the saturation of the instrument. The detuning technique enables us to derive a well calibrated image with a wide linearity range (Iwai et al. 2017).

The observed sunspot was part of the active region AR12470, located in the eastern hemisphere (N13E30) on December 16. Figure 1(a) shows the full disk radio image derived from the single-dish observation by ALMA. Figure 1(b) displays an EUV image at 1700 Å observed by the Atmospheric Imaging Assembly (AIA; Lemen et al. 2012) on board the SDO. In this paper, we use right-ascension (R.A.) and declination (Dec.) axes for image display with coordinates measuring offset from the solar disk center, Hence, images are rotated by the solar inclination angle (P=9.6°) from heliographic coordinates.

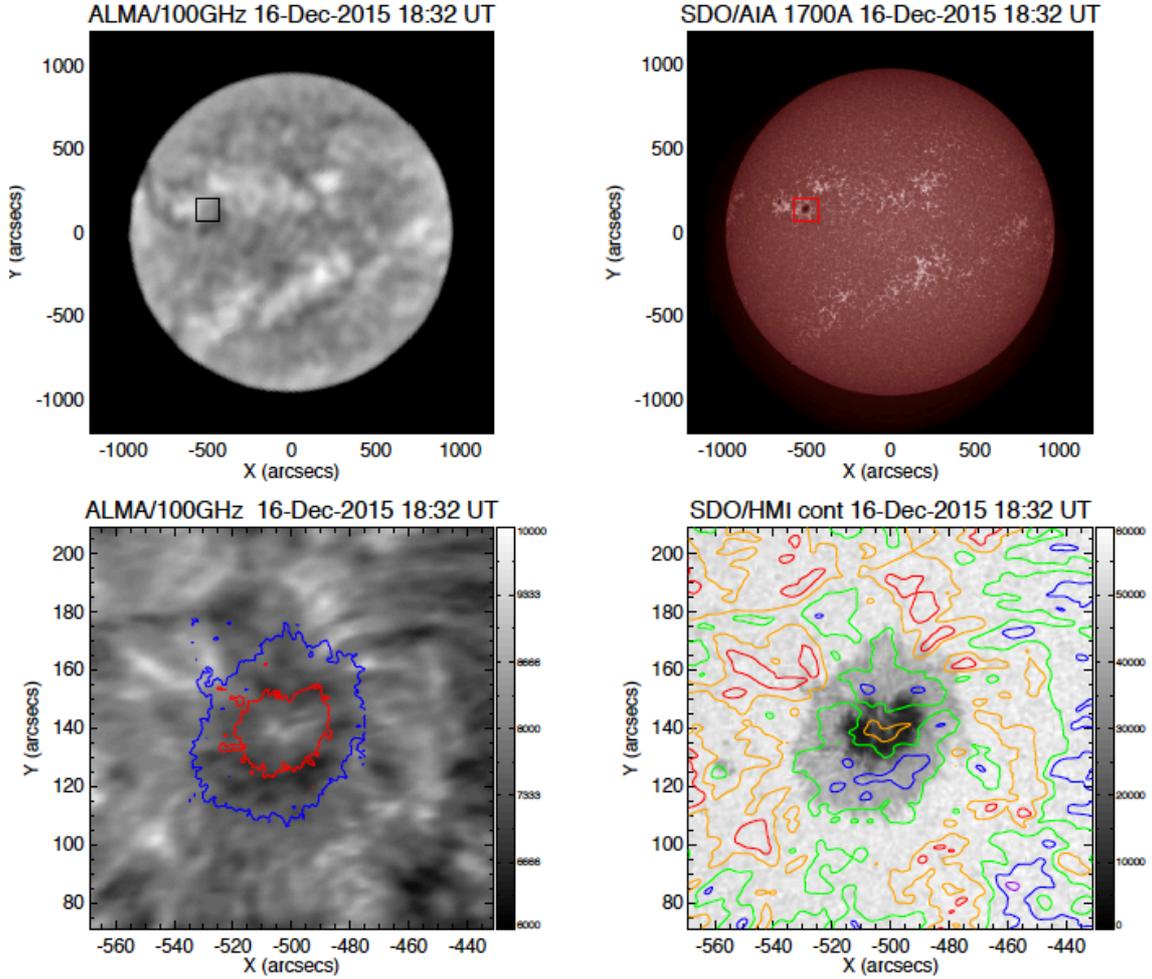

Figure 1: Full-disk solar images observed by (a) ALMA at λ=3 mm and (b) AIA at 1700 Å, on 2015 December 16 centered at 18:32 UT. Details of the region within the rectangles in panels (a) and (b), including the large sunspot of AR12470, are shown in panels (c), recorded by ALMA at 3 mm, and (d), acquired by the HMI in the visible continuum. Red and blue contours in panel (c) indicate the boundaries of the umbra (0.65 of quiet sun intensity in the visible continuum) and penumbra (0.9 of quiet sun level), respectively. The overlaid color contours in panel (d) indicate the 6300 K (purple), 6900 K (blue), 7500 K (green), 8100 K (orange), and 8700 K (red) brightness temperature levels in the ALMA λ=3 mm map. The horizontal and vertical axes are the right ascension and declination offsets, respectively, measured from the disk center position.

## 3. Data Analysis and Results

Figure 1(c) shows the ALMA λ=3 mm interferometer image of the region enclosed by the rectangles in Figures 1(a) and (b), centered on the sunspot of interest in active region AR 12470. Figure 1(d) displays the corresponding visible continuum image observed by the Helioseismic and Magnetic Imager (HMI: Scherrer et al. 2012), with temperature contours from the ALMA data overlaid. The most striking feature of Figure 1(c) is that the inner sunspot umbra is brighter than its surroundings at 3 mm, having a brightness matching the superpenumbra and plage surrounding the sunspot (compare with Figure 2(d) where the brightness contours at 3 mm are overlaid on an HMI magnetogram). The inner penumbra (7200 K) and the outer part of the umbra (7300 K) are found to be darker than the gas surrounding the sunspot and also darker than the inner umbra. Basically, the sunspot looks like a dark ring surrounding a bright inner umbra. The average temperature of this "umbral brightness enhancement" is about 8000 K, which is about 800 K higher than the surrounding inner penumbral region and 400 K higher than the outer penumbra. The plage region neighboring the sunspot is about 7800 K at 3 mm. The noise level of this map is about 3.7 K (Shimojo et al., 2017) and is not significant relative to the magnitude of the temperature differences that we discuss.[1] Note that the spatial resolution of the full-disk image in Figure 1(a) is about 60". Hence, the small-scale brightness distributions in Figure 1(c) are smoothed over in the full-disk image.

Figure 2 displays the comparison between radio and EUV images at 1700 (lower photosphere), 304 (transition region), and 171 Å (cool corona) observed by AIA and the line-of-sight magnetic field observed by HMI. The 1700 Å image shows a dark sunspot cooler than the surrounding photosphere, and the radio brightness temperature at 3 mm generally correlates well with the 1700 Å emission. The center of the umbra, however, is brighter than its surroundings in the λ=3 mm map, whereas it is the darkest region in the 1700 Å map.

---

[1] Note that the quiet-Sun disk center temperature of the single-dish image used here is around 7500 K, i.e., 200 K larger than the value recommended for scaling by White et al. (2017). Since in this paper we are only concerned with relative temperature differences, which are not affected by the absolute scale, we choose not to correct the absolute temperatures for the 2.7% offset represented by 200 K. It should also be mentioned that we combined the visibilities derived from the interferometer and single dish using the default parameters, which might cause an additional offset (Shimojo et al. 2017). This offset also has no influence for our discussions.

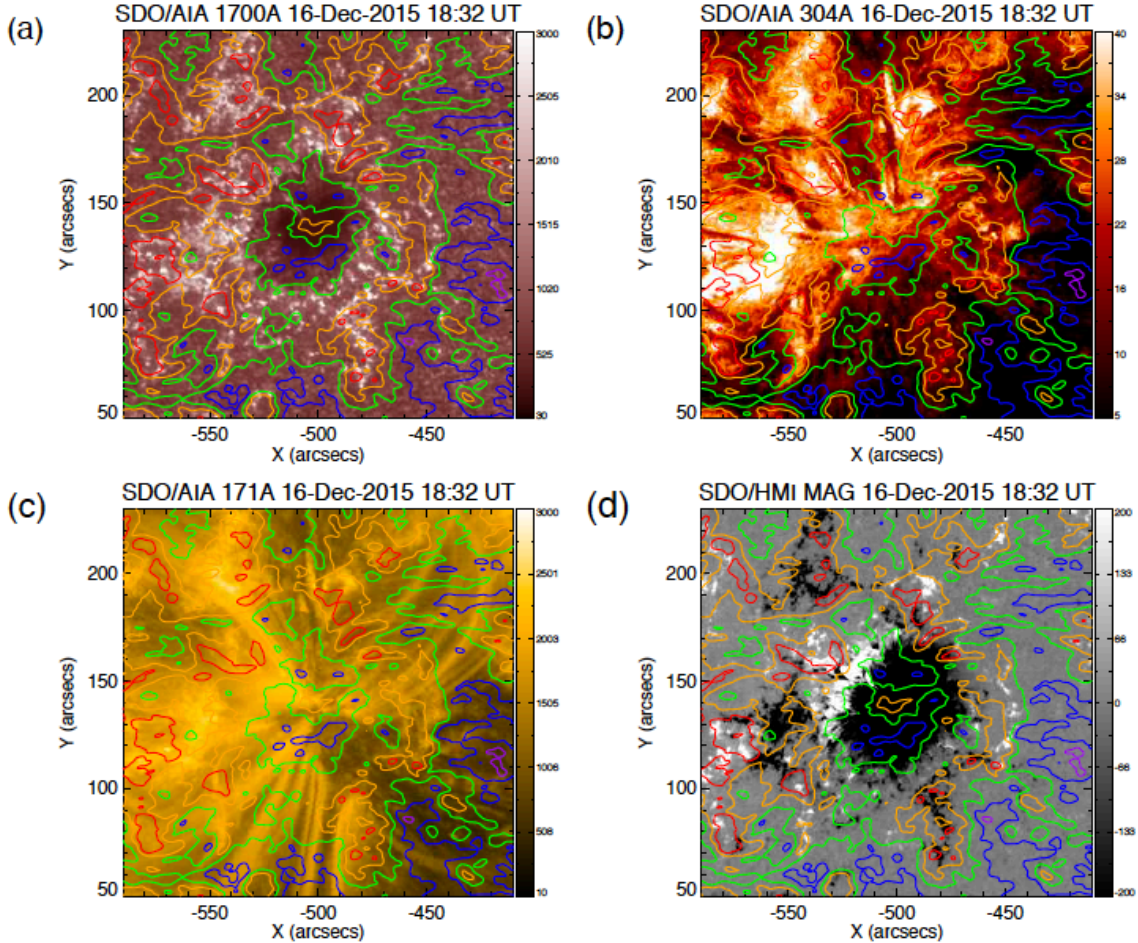

Figure 2: EUV images of the sunspot indicated by the rectangle in Figure 1(a), at (a) 1700 Å, (b) He II 304 Å, and (c) Fe IX 171 Å acquired by the AIA on 2015 December 16, and (d) line-of-sight magnetogram obtained by HMI. The overlaid contours indicate the radio brightness temperature at λ=3 mm observed by ALMA at the same brightness temperatures as in Figure 1(d).

The λ=3 mm and 304 Å images also show a general similarity outside the sunspot, which is difficult to identify in the 304 Å image where the leading side of the spot is dark and the trailing side is bright. The agreement between the λ=3 mm and coronal 171 Å images is weaker still. However, the 171 Å image does indicate that a number of bright loops have footpoints in either the umbra or penumbra. The umbral core appears to harbor the footpoints of a bright set of coronal loops.

In Figure 3 the brightness contours of λ=3 mm radiation are overlaid on slit-jaw images at 1330 Å (C II), 1400 Å (Si IV), 2796 Å (Mg II line core) and 2832 Å (Mg II line wing, looking deeper into the atmosphere than the line core) obtained by IRIS (De Pontieu et al., 2014), quasi-simultaneously. The λ=3 mm image looks similar to the 1330 Å and 1400 Å images, which are thought to be emitted from the upper chromosphere and lower transition region, respectively. There is a weak brightening in the 1330 and 1400 Å images at the site of the millimeter umbral brightness enhancement surrounded by dark penumbra, as can be seen in Figures 3(a) and 3(b). In the UV images this feature lies at the end of a much brighter intrusion into the umbra from the northeast that appears to be a partial lightbridge.

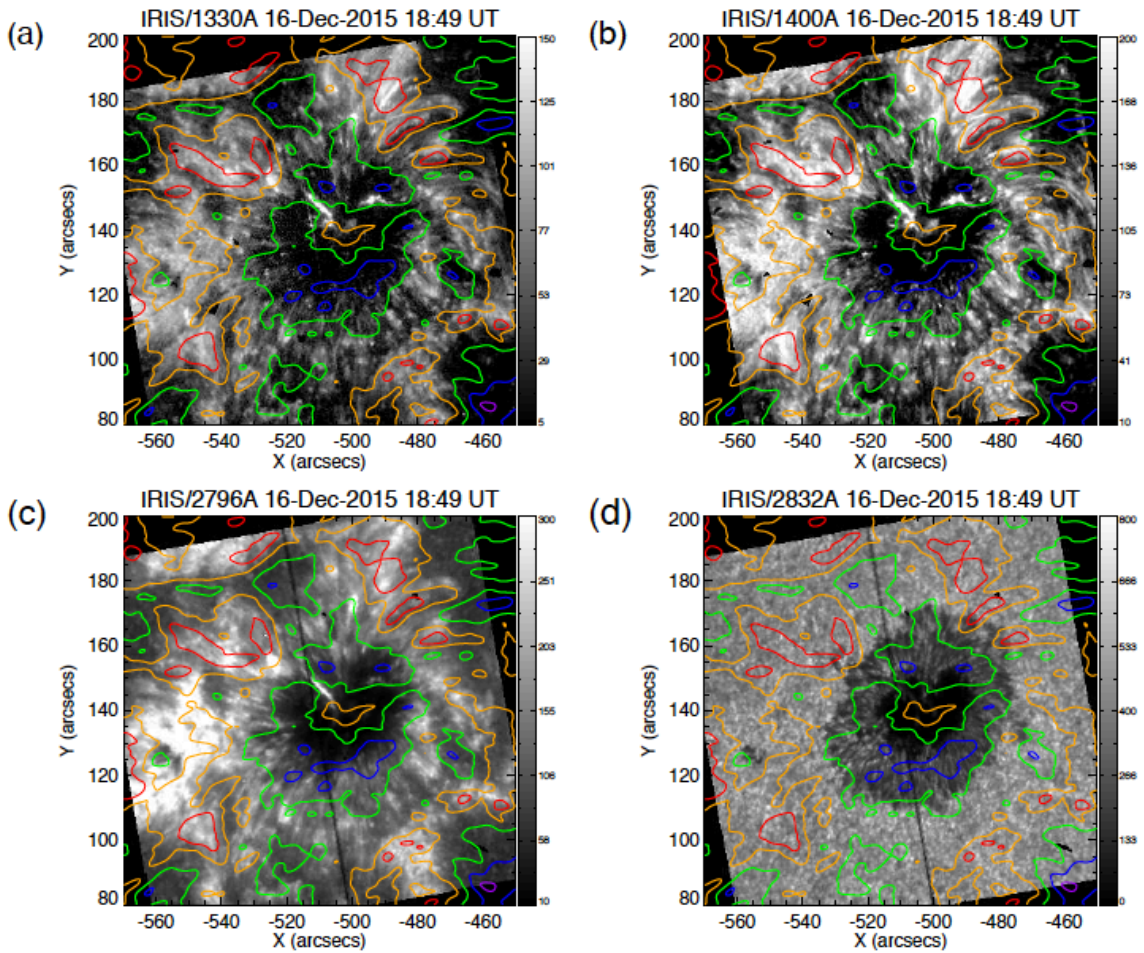

Figure 3: IRIS slit-jaw images of the region within the rectangle in Figure 1(a) at (a) 1330 Å (b) 1400 Å, (c) 2796 Å, and (d) 2832 Å, acquired on 2015 December 16. The overlaid color contours indicate the radio brightness temperature at 3 mm observed by ALMA, corresponding to the same brightness temperatures as in Figure 1(d).

Figure 4 compares a sequence of 1600 Å images taken by AIA during the ALMA observation with contours obtained from ALMA. The 1600 Å bandpass of AIA includes the CIV line originating in the transition region (Lemen et al. 2012) and so should have a response similar to the IRIS 1400 Å line. There is a small-scale brightening in the sunspot at 1600 Å around the inner part of the partial lightbridge, which, however, is not co-spatial with the main brightening seen in ALMA band 3. While the lightbridge shows variable brightness in this 1600 Å sequence, no variability is seen in the center of the umbra.

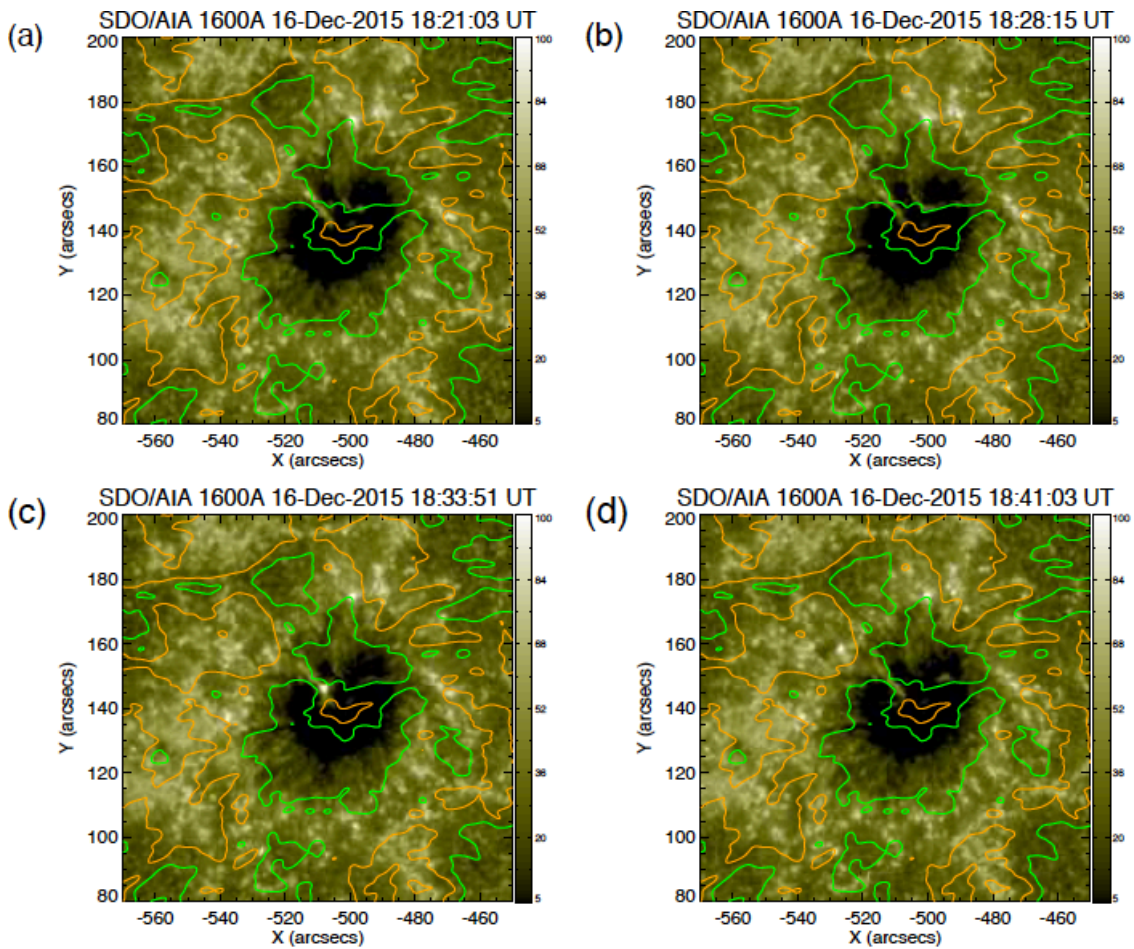

Figure 4: 1600 Å images observed by AIA at (a) 18:23, (b) 18:28, (c) 18:33, and (d) 18:41on 2015 December 16. 1600 Å images in (a) and (d) are shifted to cancel the solar differential rotation. The overlaid color contours indicate the radio brightness temperature at 3 mm observed by ALMA at 18:32 (see Figure 1(d)).

The right panels of Figure 5 show histograms of the radio brightness temperature,

visible continuum intensity, and line-of-sight magnetic field derived from different umbral regions. The red histograms correspond to the inner umbral regions enclosed by the red ellipse in the left panels, while blue histograms refer to the outer umbra lying outside the blue ellipses and inside the green lines which indicate the umbral boundary. Panel (b) strikingly emphasizes the temperature difference between the warm inner umbra and the cool outer umbra at $\lambda=3$ mm: to our knowledge, there are no models predicting such behavior. The brightness distributions are almost reversed between the radio and visible continuum emission (panel (d)). The magnetic field is, on average, stronger within the red ellipse, although there are also strong field values outside that region (panel (f)). The general behavior of the two lower histograms agrees with the anti-correlation between brightness and field strength that is well known for sunspot umbrae (e.g., Martinez Pillet et al. 1993; Solanki et al. 1994; Tiwari et al. 2015).

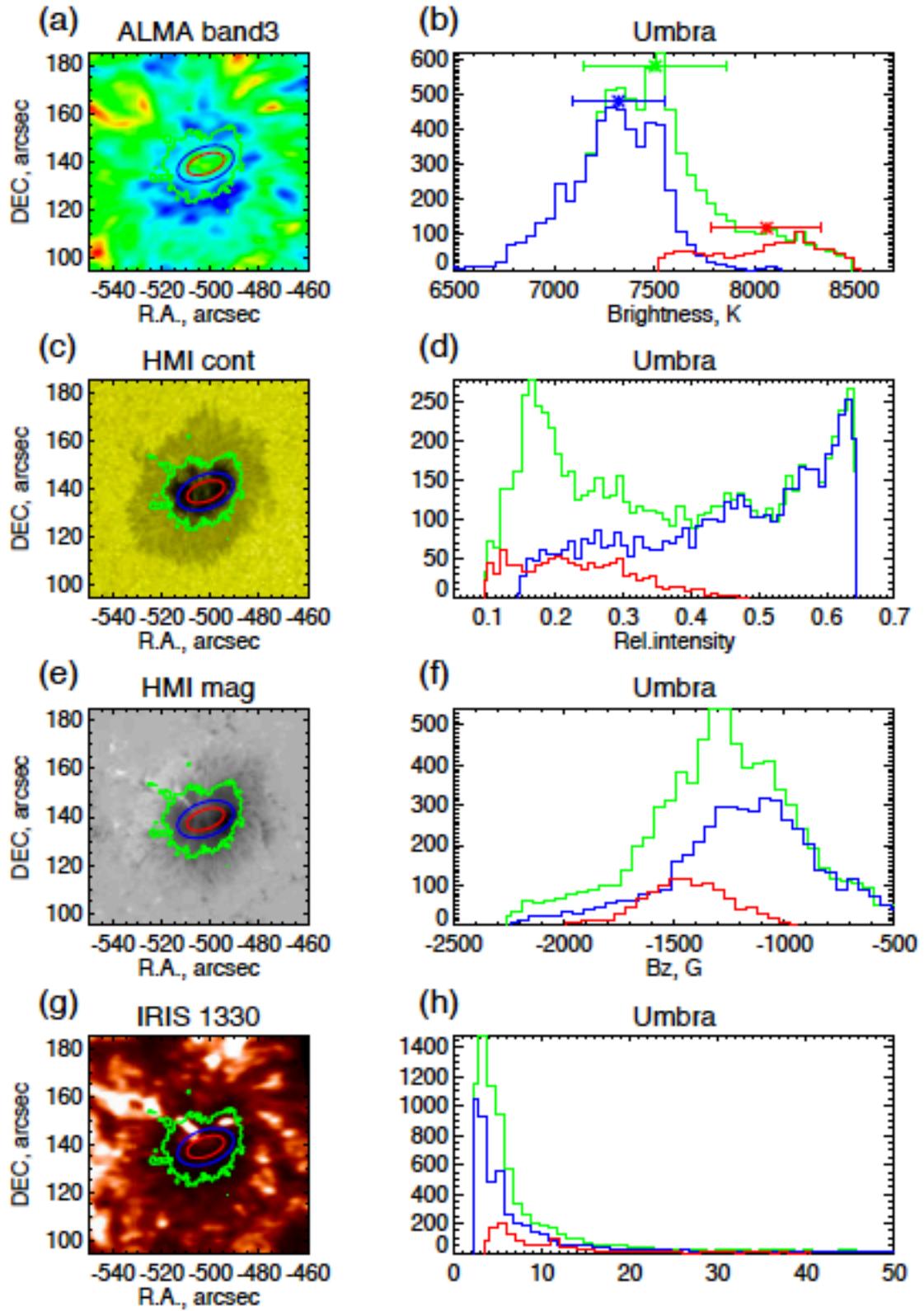

Figure 5 Left panels: Images of (a) λ=3 mm brightness temperature, (c) visible continuum intensity, (e) line-of-sight magnetic field, and (g) 1330 Å convolved to the ALMA beam size. Right panels: Histograms from the images of (b) brightness temperature at λ=3 mm, (d) visible continuum emission, (f) strength of the line-of-sight magnetic field, and (h) 1330 Å, derived from different umbral regions. Ellipses in the left panels outline regions over which the histograms in the right panels were made. Green histograms: whole umbra within green contour, red histogram: inner umbra within red ellipse, and blue histogram: outer umbra between blue ellipse and green contour. In the top panel asterisks and error bars indicate average values and their standard deviations for each region, respectively.

## 4 Summary and Discussion

### 4.1 Observational summary

We have analyzed an ALMA Band 3 mosaic image of a large sunspot. The map, with 4".9×2".2 spatial resolution, represents the highest spatial resolution λ=3 mm map of an entire sunspot yet available. We also compared the radio image with UV, EUV and visible wavelength images observed by IRIS, AIA and HMI. The sunspot is not immediately obvious in the ALMA image (Fig. 1c), at least in part because the surprising umbral brightening disrupts what would otherwise be a dark bowl coincident with the sunspot, leading instead to a dark ring over the outer umbra. The outer penumbra is nearly as bright as the surrounding active-region plage at λ=3 mm.

The λ=3 mm umbral brightness enhancement makes the umbral core nearly as bright as the surrounding plage. The IRIS observations revealed a bright region at the center of the umbra in 1330 and 1400 Å images (Figure 3(a) and (b)), which is close to but not identical with the millimetric umbral brightness enhancement. However, there was no clear counterpart of the millimetric umbral brightness enhancement in the 1700, 304, or 171 Å images, which are otherwise generally similar to the λ=3 mm image in the emission from bright plage surrounding the sunspot (Figs. 2, 3). There was a partial lightbridge intruding into the upper left part of the umbra (Figs. 1(d), 2(d)) that was bright in the 1700, 1330 and 1400 Å images from the upper chromosphere (Figs. 2(a), 3(a), 3(b)). The radio umbral brightness enhancement was located adjacent to the inner boundary of the lightbridge, but it is striking that the lightbridge itself is less apparent in the millimeter image.

As shown in Figure 5(g), the pattern of bright and dark structures seen in the IRIS 1330

Å image looks very similar to the λ=3 mm image even after degrading to ALMA resolution. However, Figure 5(h) shows that, unlike the λ=3 mm image, the 1330 Å histogram does not show a clear distinction between the inner and outer penumbra. Hence the λ=3 mm radiation is providing a view of the umbra distinctly different from that of other chromospheric diagnostics.

### 4.2. Bright region in the umbral core

Although there have been several sunspot observations at λ=3 mm (e.g., White et al. 2005, Iwai and Shimojo. 2015), a bright structure at the umbral center, as found here, has never previously been reported. The spatial scale of the observed umbra and corresponding radio-bright structure is about 10", which is sufficiently large to be resolved by the beam of ALMA in this study (4".9×2"2). However, the spatial resolutions of previous observations would not have resolved this structure. Hence, it is not surprising that ALMA is the first instrument to discover such an umbral brightness enhancement.

### 4.3 Origin of the umbral brightness enhancement

There are several possible explanations for the enhanced radio brightness in the central umbra. The first is that it is an intrinsic property of sunspots. The observed umbral brightness temperature was around 900 K higher than cool quiet-Sun regions in the ALMA image, which is in fact consistent with the umbral brightening relative to the quiet Sun suggested by a number of existing models of the umbral atmosphere (e.g., see Loukitcheva et al 2014). This suggests that the discovery in this observation might not be of an "umbral brightness enhancement" but rather of a "penumbral darkening". ALMA data will be used to test sunspot atmosphere models in a follow-up paper. Note that the ALMA science verification data contains 3 mm images of the same active region observed on the following day (December 17, 2015). Those images also show a small scale bright structure above the umbra, which suggests that the observed brightness enhancement in this study was not a singular event.

In this case, the findings from Figure 5 suggest that the umbral magnetic field, which inhibits convective energy transport in the photosphere and thus leads to a darkening in those layers, reverses its influence and produces an enhanced emission layer at λ=3 mm, likely in the upper chromosphere. This is not entirely a 1-to-1 relationship, as there are also very cool regions in the photosphere, located between the blue and red ellipses, that are not particularly bright in the chromospheric λ=3 mm radiation.

Another possibility is that the λ=3 mm umbral brightness enhancement is related to a coronal plume. In Figure 2(c), the region of enhanced millimetric brightness corresponds to footpoints of coronal loops at 171 Å inside the umbra. In addition, there is a region bright at 1330 and 1400 Å at the center of the umbra (see Fig. 3(c,d),). These characteristics are associated with coronal plumes (see, e.g., Tian et al. 2009), which have been seen in longer-wavelength radio observations as well (Brosius & White 2004). It should also be mentioned that a downflow is usually observed in chromospheric lines in plumes (Straus et al. 2015; Chitta et al. 2016), which might be associated with coronal rain or a siphon flow. The downflowing coronal material interacting with the dense lower atmosphere could result in heating and enhancement of the radio emission.

Three-minute oscillations originating in the photosphere are usually observed in umbral regions in many chromospheric diagnostics, including at millimeter wavelengths (e.g. Loukitcheva et al. 2006). These waves tend to form shocks in the solar chromosphere which result in brightenings in the cores of chromospheric spectral lines such as Hα and the Ca II IR triplet, referred to as "umbral flashes". These "flashes" of enhanced emission typically last less than a minute. Using a sophisticated and computationally intensive non-LTE inversion technique, de la Cruz Rodriguez et al (2013) inferred a temperature contrast of order 1000 K in umbral flashes at the level of the temperature minimum (3000 K temperature level) from fits to spectrally- and spatially-resolved measurements of the Ca II 854.2 nm line. The temperature enhancement at greater heights was somewhat smaller in their results, but shock formation is still expected at the height of the ALMA emission (7500 K). Hence, umbral flashes are a possible explanation for the radio brightness enhancement found in this study. We stress, however, that the ALMA data provide measurements of the brightness temperature enhancement without the need for any spectral inversions and without invoking non-LTE, demonstrating the advantages of radio observations. Umbral flashes are intrinsically time variable, providing a test for future observations.

In summary, the chromosphere of the sunspot studied here displayed an unexpected structure at millimeter wavelengths, with a cool inner penumbra (and outer umbra), but a hot inner umbra. The present observations cannot distinguish between different scenarios for the brightness enhancement, so that additional millimeter observations are needed to understand umbral brightness. In particular, a time series of the umbra will be able to uncover possible time variability of the umbral brightness enhancement, which will be essential to distinguish between the different possible sources discussed here. Simultaneous observations at multiple radio frequency bands will help to understand the height distribution of this phenomenon. Finally, the initial science verification data used in this study have poorer spatial resolution than the ALMA solar data expected from full science observations, which will be of order 1 arcsec at $\lambda$=3 mm, and we expect that further umbral chromospheric sub-structure will be visible with the better resolution.

Acknowledgements
We thank Takenori J Okamoto for his help the IRIS data analysis. This paper makes use of the following ALMA data: ADS/JAO.ALMA#2011.0.00020.SV. ALMA is a partnership of ESO (representing its member states), NSF (USA) and NINS (Japan), together with NRC (Canada) and NSC and ASIAA (Taiwan), and KASI (Republic of Korea), in cooperation with the Republic of Chile. The Joint ALMA Observatory is operated by ESO, AUI/NRAO and NAOJ. The AIA and HMI data are courtesy of the NASA/SDO, as well as AIA and HMI science teams. IRIS is a NASA small explorer mission developed and operated by LMSAL with mission operations executed at NASA Ames Research center and major contributions to downlink communications funded by ESA and the Norwegian Space Centre. KI is supported by a Japan Society for the Promotion of Science (JSPS) Research Fellowship. ML acknowledges NSF grant AST-1312802, NASA grant NNX14AK66G, Russian RFBR grants 15-02-03835 and 16-02-00749, and Saint-Petersburg State University grant 6.37.343.2015.

References
Bastian, T. S., Ewell, M. W., Jr., & Zirin, H. 1993, ApJ, 415, 364
Brosius, J. W., & White, S. M. 2004, ApJ 601, 546
**Chitta, L. P., Peter, H., & Young, P. R. 2016, A&A, 587, A20**
**de la Cruz Rodríguez, J., Rouppe van der Voort, L., Socas-Navarro, H., &**
**van Noort, M. 2013, A&A, 556, A115**
De Pontieu, B., Title, A. M., Lemen, J. R., et al. 2014, SoPh, 289, 2733
Dulk, G. A. 1985, ARA&A, 23, 169
Hills, R.E., Kurz, R.J., Peck, A.B., 2010, Proceedings of the SPIE 7733, 773317
Iwai, K., & Shimojo, M. 2015, ApJ, 804, 48
Iwai, K., Koshiishi, H., Shibasaki, K., et al. 2016, ApJ, 816,91
Iwai, K., Shimojo, M., Asayama, S., et al. 2017, SoPh, 292:22

Lemen, J. R., Title, A. M., Akin, D. J., et al. 2012, SoPh, 275, 17

Lindsey, C., & Kopp, G. 1995, ApJ, 453, 517

Loukitcheva, M., Solanki, S. K., White, S. 2006, A&A, 456, 713

Loukitcheva, M., Solanki, S. K., & White, S. M. 2014, A&A, 561, 133

Loukitcheva, M., Solanki, S. K., Carlsson, M., White, S. M. 2015, A&A, 575, 15

Martinez Pillet, V., Vazquez, M. 1993, A&A, 270, 494

Rempel, M., & Schlichenmaier, R. 2011, LRSP, 8, 3

Scherrer, P. H., Schou, J., Bush, R. I. et al. 2012, SoPh, 275, 207

Shimojo, M., Bastian, T. S., Hales, A. S., et al. 2017, Solar Physics, submitted

Solanki, S. K., Montavon, C. A. P., Livingston, W. 1994, A&A, 283, 221

Solanki, S. K. 2003, A&ARv, 11, 153

Straus, T., Fleck, B., & Andretta, V. 2015, A&A, 582, A116

Tian, H., Curdt, W., Teriaca, L., Landi, E., & Marsch, E. 2009, A&A, 505, 307

Tiwari, S. K., van Noort, M., Solanki, S. K., & Lagg, A. 2015, A&A, 583, A119

Vernazza, J. E., Reeves, E. M. 1978, ApJS, 37, 485

White, S., Loukitcheva, M., & Solanki, S. K. 2006, A&A, 456, 697

White, S. M., Iwai, K., Phillips, N. M. et al. 2017, 2017, Solar Physics, submitted